\documentclass{elsart}
\usepackage{amsmath}
\usepackage{epsfig}
\usepackage{hyperref}

\usepackage{graphicx}
\usepackage[usenames]{color}
\usepackage[normalem]{ulem}

\begin{document}
\begin{frontmatter}

\title{Collisional Dynamics of Solitons in the Coupled $\cal{PT}$ Symmetric Nonlocal Nonlinear Schr\"odinger Equations}

\author{P. S. Vinayagam$^{1}$}
\author{R. Radha$^{2}$}
\author{U. Al Khawaja$^{\ast,^1}$}
\corauth{Corresponding author.} \ead{u.alkhawaja@uaeu.ac.ae}
\author{Liming Ling$^{3}$}
\address{$^1$ Department of Physics,
United Arab Emirates University, P.O.Box 15551, Al-Ain, United
Arab Emirates \\
$^2$ Centre for Nonlinear Science (CeNSc), PG and Research
Department of Physics, Government College for Women (Autonomous),
Kumbakonam 612001, India\\
$^{3}$Department of Mathematics, South China University of
Technology, Guangzhou 510640, China}

\begin{abstract}
We investigate the focussing coupled $\cal{PT}$ symmetric nonlocal
nonlinear Schr\"odinger equation employing Darboux transformation
approach. We find a family of exact solutions including pairs of
Bright-Bright, Dark-Dark and Bright-Dark solitons in addition to
solitary waves. We show that one can convert bright bound state
onto a dark bound state in a two soliton solution by selectively
finetuning the amplitude dependent parameter. We also show that
the energy in each mode remains conserved unlike the celebrated
Manakov model. We also characterize the behaviour of the soliton
solutions in detail. We emphasize that the above phenomenon occurs
due to the nonlocality of the model.
\end{abstract}
\end{frontmatter}
\newpage

\section{Introduction}
Solitons are an important subject of study in both theoretical and
experimental aspects in various fields, such as nonlinear optics,
Bose-Einstein condensates, plasma physics and hydrodynamics.
Nonlinear differential equation which describes these physical
media gives rise to solitonic solutions. In the study of nonlinear
wave propagation, by generating explicit soliton solutions,
\textit{exactly solvable models} play a pivotal role. There are
many physically significant integrable systems which apply to
diverse problems in nonlinear optics. In particular, the nonlinear
Schr\"odinger equation (NLSE) and its coupled equation (CNLSE)
have been most thoroughly investigated as integrable models,
especially in nonlinear optics, where many species of solitons
have been predicted \cite{opticalsoliton}. Finding new integrable
reductions of known nonlinear differential equations like NLSE or
CNLSE is a new trend in the theory of integrable systems. This is
motivated by the fact that, new integrable systems may give rise
to new solutions. This, inturn,  provides new ideas to proceed
further  to unearth significant results in future, and also open
new avenues in the respective fields.

As mentioned above, recently Ablowitz and Musslimani proposed a
new nonlocal NLSE \cite{ablo-prl}
\begin{align}\label{ablowitz model}
\textit{i} q_t(x,t)= q_{xx}(x,t)+2 g q(x,t)q^{\ast}(-x,t)q(x,t),
\end{align}
where, $q(x,t)$ is a complex valued function of real variables $x$
and $t$, and $g=\pm 1$ denotes, respectively, the focussing (+)
and defocussing (-) cases. The above equation (\ref{ablowitz
model}) is $\cal{PT}$ symmetric in the sense that the equation
brings a self induced potential of the form
$V(x,t)=q(x,t)q^{\ast}(-x,t)$ and satisfies the $\cal{PT}$
symmetric condition $V(x,t)=V^{\ast}(-x,t)$. In general,
$\cal{PT}$ symmetric nonlinear Schr\"odinger systems  should have
linear potential term of the following form, namely
\begin{align}
\textit{i} q_t(x,t)+q_{xx}(x,t)+V(x,t)q(x,t)+2\lvert
q(x,t)\rvert^2 q(x,t)=0.
\end{align}
This equation has been explored well in the last few years. For
example: it includes the effect of nonlinearity on beam dynamics
in $\cal{PT}$ symmetric potential \cite{exp1}, solitons in
dual-core waveguides \cite{exp2,exp3}, and Bragg solitons in
$\cal{PT}$ symmetric potentials \cite{exp4}. But, Ablowitz and
Musslimani found the alternative way to fix the integrability in a
nonlocal Schr\"odinger type equation by replacing the standard
nonlinearity $\lvert q \rvert^2 q$ with its $\cal{PT}$ symmetric
counterpart $q(x,t)q^{\ast}(-x,t)q(x,t)$. Interestingly, in this
study Ablowitz \textit{et al.,} have shown the new model given by
Eq. \eqref{ablowitz model} is fully \textit{integrable} since it
possesses linear Lax pair and infinite number of conserved
quantities and \textit{ nonlocal} in the sense that the evolution
of the field at transverse coordinate $x$ always requires the
information from the opposite point $-x$ \cite{ablo-prl} .

As we mentioned before, new model reduction to the well known
models provides us a chance to unearth new results like, the above
new integrable model admits both bright and dark soliton solutions
for the same nonlinear interaction $g<0$ with positive dispersion
\cite{aksarma}. It is in contrast with the usual local NLSE, since
it admits only bright and dark soltions for attractive ($g>0$) and
repulsive ($g<0$) nonlinear interaction, respectively with
positive dispersion. Realization of nonlocal nonlinearity is
witnessed in many experiments such as, in diffusion of charge
carriers, atoms, or molecules in atomic vapors
\cite{realexp1,realexp2}. Also, it is witnessed in the study of
BEC with a long-range interaction. The BEC with magnetic
dipole-dipole forces was reported in \cite{realexp3} and occurred
in the case of optical spatial solitons in a highly nonlocal
medium \cite{realexp4}. It has been shown that nonlocal $\cal{PT}$
symmetric integrable equations can provide an ideal test bed for
researchers to study and observe the ramification of nonlocality
in optics. Due to the fact that, in optics, the paraxial equation
of diffraction is mathematically isomorphic to the Schr\"odinger
equation in quantum mechanics
\cite{optics1,optics2,optics3,optics4}, this analogy allowed
observation of $\cal{PT}$ symmetry in optical waveguide structures
and lattices \cite{exp4,optics5,optics6}. This study of $\cal{PT}$
symmetric concepts in optics can also lead to alternative classes
of optical structures and devices such as, non-Hermitian Bloch
oscillations \cite{bloch}, simultaneous lasing-absorbing
\cite{lasing,absorbing}, and selective lasing
\cite{select-lasing}. Finally, $\cal{PT}$-symmetric concepts have
also been used in plasmonics \cite{plasmonics}, optical
metamaterial \cite{opt-meta-1,opt-meta-2} and coherent atomic
medium \cite{cohe-atom}.

Motivated by the above significant results and unique dynamical
behavior, we investigate the $\cal{PT}$-symmetric coupled nonlocal
nonlinear Schr\"odinger equation (PTCNNLSE) in continuous media.
Our analysis indicates that the $\cal{PT}$ symmetric CNLSE
exhibits unique dynamics and features that one can convert bright
bound state onto a dark bound state in a two soliton solution by
selectively finetuning the amplitude-dependent parameter while the
energy in each mode remains conserved unlike the Manakov model.

The plan of the paper is as follows. In section II, we present the
mathematical (integrable) model governing the dynamics of nonlocal
$\cal{PT}$ symmetric coupled nonlinear Schr\"odinger equation and
its Lax pair and then derive the explicit soliton solution. In
section III, we present the family of all solitonic solutions
generated from the general solution obtained in section II. In
section IV, we discuss the collisional dynamics of solitons and
the mechanism which involves the conversion of bright to dark
bound state. The results are then summarized in section V.

\begin{figure*}[!ht]
\centering\includegraphics[width=1.0\linewidth]{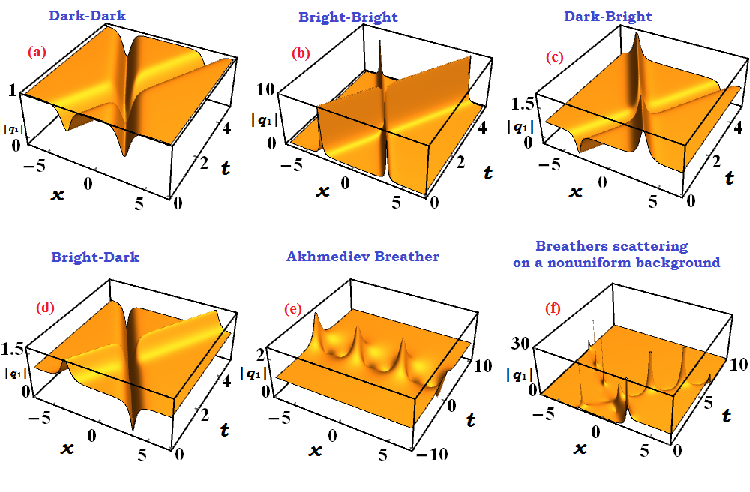}
\caption{Families of exact two soliton solutions for the choice of
parameter $\alpha_1=1, \alpha_2=0.5, a=-4, c=-1, b=0, d=0, a_1=8,
\zeta_1=0, \eta_1=0.25\pi$ and manipulation of amplitude dependent
parameter $b_1$ leads to cases (a) dark-dark for $b_1=7$, (b)
bright-bright for $b_1=3.9$, (c) dark-bright for $b_1=2.8$, (d)
bright-dark for $b_1=5$ and special cases like oscillatory
solutions, namely (e) Akhmediev breather  and (f) the difference
between the background levels for the choice of parameter
selectively chosen to achieve $J_1=-1 (<0)$)}\label{grid}
\end{figure*}

\section{Coupled $\cal{PT}$ symmetric nonlocal NLSE}
Let us consider, the coupled $\cal{PT}$ symmetric nonlocal NLSE of
the following form,
\begin{subequations}\label{eq:ptcnnls}
\begin{align}
    \textit{i}\frac{\partial q_{1}(x,t)}{\partial t}+\frac{1}{2}\frac{\partial^2 q_{1}(x,t)}{\partial x^2}+[q_1(x,t)q_1^{\ast}(-x,t)+q_2(x,t)q_2^{\ast}(-x,t)]q_1(x,t)&=0,\label{ptcnnls1}\\
    \textit{i}\frac{\partial q_{2}(x,t)}{\partial t}+\frac{1}{2}\frac{\partial^2 q_{2}(x,t)}{\partial x^2}+[q_1(x,t)q_1^{\ast}(-x,t)+q_2(x,t)q_2^{\ast}(-x,t)]q_2(x,t)&=0,\label{ptcnnls2}
\end{align}
\end{subequations}
where, $q_{j}(x,t)\, (j=1,2)$ are the two complex field
amplitudes. This equation is also a non-Hermitian $\cal{PT}$
symmetric system in the sense that the self-induced potential
$V(x,t)= q_1(x,t)q_1^{\ast}(-x,t)+q_2(x,t)q_2^{\ast}(-x,t)$
satisfies the $\cal{PT}$ symmetric condition
$V(x,t)=V^{\ast}(-x,t)$. It is worth mentioning here, that
Eq.~\eqref{eq:ptcnnls} is nonlocal, i.e., the evolution of the
field variables $(q_{1}(x,t), q_{2}(x,t))$ at the transverse
coordinate $x$ always requires information from the opposite point
$-x$. The above coupled Eqs.~(\ref{ptcnnls1}) and (\ref{ptcnnls2})
admit the following Lax pair
\begin{subequations}%
\begin{align}%
\Phi_{x} +{\bf U}\Phi =& 0,  \label{Phix} \\
\Phi_{t} +{\bf V}\Phi =& 0,  \label{Phit}
\end{align}
\end{subequations}%
where $\Phi =(\phi_{1},\phi _{2},\phi_{3})^{T}$ is a
three-component Jost function,
\begin{align}
{\bf U}=\begin{pmatrix}
2 i\lambda & i q_{1}(x,t) & i q_{2}(x,t) \\
i q^{\ast}_{1}(x,t) & 0 & 0 \\
i q^{\ast}_{2}(x,t) & 0 & 0%
\end{pmatrix},
\;\;\; {\bf V}=
\begin{pmatrix}
{\bf V}_{11} & {\bf V}_{12} & {\bf V}_{13} \\
{\bf V}_{21} & {\bf V}_{22} & {\bf V}_{23} \\
{\bf V}_{31} & {\bf V}_{32} & {\bf V}_{33}%
\end{pmatrix},
\label{UV}
\end{align}
with
\begin{align}
{\bf V}_{11} = &\textit{i} \lambda^2-\frac{1}{2}\textit{i}\left[
q_{1}(x,t)q^{\ast}_{1}(x,t)+q_{2}(x,t)q^{\ast}_{2}(x,t) \right], \notag \\
\notag \\
{\bf V}_{12} = &\textit{i} \lambda q_{1}(x,t) +
\frac{1}{2}q^{\ast}_{1x}(x,t),\,\,
{\bf V}_{13} = \textit{i} \lambda q_{2}(x,t) + \frac{1}{2}q^{\ast}_{2x}(x,t),\notag\\
{\bf V}_{21} = & \textit{i} \lambda q^{\ast}_{1}(x,t) - \frac{1}{2}q^{\ast}_{1x}(x,t),  \notag \\
{\bf V}_{22} = &-\textit{i} \lambda^2+\frac{1}{2}\textit{i}\left[
q_{1}(x,t)q^{\ast}_{1}(x,t)\right], \notag \\
{\bf V}_{23} = & -\frac{\textit{i}}{2} q_{1}(x,t)q^{\ast}_{1}(x,t),  \notag \\
{\bf V}_{31} = & \textit{i} \lambda
q^{\ast}_{2}(x,t)-\frac{1}{2}q^{\ast}_{2x}(x,t),\,\
{\bf V}_{32} = -\frac{\textit{i}}{2} q_{2}(x,t)q^{\ast}_{2}(x,t),\notag\\
{\bf V}_{32} = &-\frac{\textit{i}}{2} q_{2}(x,t)q^{\ast}_{2}(x,t),\notag \\
{\bf V}_{33} = &-\textit{i} \lambda^2+\frac{1}{2}\textit{i}\left[
q_{2}(x,t)q^{\ast}_{2}(x,t)\right], \notag
\end{align}%
where $\lambda$ is the spectral parameter. The compatibility
condition (or the Zero Curvature condition)$ {\bf U}_{t}-{\bf
V}_{x}+[{\bf U},{\bf V}]=0$ generates the classical local coupled
NLS equation of the following form
\begin{eqnarray}
    \textit{i}\frac{\partial q_{1}}{\partial t}+\frac{1}{2}\frac{\partial^2 q_{1}}{\partial x^2}+[q_1(x,t)q_1^{\ast}(x,t)+q_2(x,t)q_2^{\ast}(x,t)]q_1(x,t)&&=0,\\
    \textit{i}\frac{\partial q_{2}}{\partial t}+\frac{1}{2}\frac{\partial^2 q_{2}}{\partial
    x^2}+[q_1(x,t)q_1^{\ast}(x,t)+q_2(x,t)q_2^{\ast}(x,t)]q_2(x,t)&&=0.
\end{eqnarray}
We then employ the strategy  proposed by Ablowitz and Mussilimani
\cite{ablo-prl} and sustain the integrability (or the lax-pair) by
replacing the nonlinear interaction term $[|q_1|^2 +|q_2|^2] q_1$
and $[|q_1|^2 +|q_2|^2] q_2$ in the above coupled equations by
their $\cal{PT}$ symmetric counterparts, namely, $[q_{1}
q_{1}^{\ast}(-x,t) +q_{2} q_{2}^{\ast}(-x,t)] q_1$ and $[q_{1}
q_{1}^{\ast}(-x,t) +q_{2} q_{2}^{\ast}(-x,t)] q_2$, to obtain the
${\cal PT}$ symmetric coupled nonlocal NLSE given by
Eqs.~(\ref{ptcnnls1}) and (\ref{ptcnnls2}).

As mentioned in the introduction, this new integrable PTCNNLSE
equation admits both bright and dark solitons for the same
nonlinearity unlike the conventional Manakov model (CNLSE). The
nonsingular bright soliton solution of Eq.\eqref{eq:ptcnnls}, of
the following form (was constructed by following the details given
by Liming et al.,\cite{liming})

\begin{subequations}\label{brightone}
    \begin{align}
       q_1^{(1)}=&2a \sin(\theta)\mathrm{sech}(2ax+\textit{i}k){\rm e}^{2\textit{i}a^2t},  \\
       q_2^{(1)}=&2a \cos(\theta)\mathrm{sech}(2ax+\textit{i}k){\rm e}^{2\textit{i}a^2t},
    \end{align}
\end{subequations}
where $a,k,\theta\in \Re$, $k\neq\pi$. The nonsingular single dark
solutions for Eqs. \eqref{eq:ptcnnls} are
\begin{subequations}\label{darkone}
    \begin{align}
       q_1^{(1)}=&a\sin(\theta)\tanh(ax+\textit{i}k){\rm e}^{-\textit{i}a^2t},  \\
       q_2^{(1)}=&a\cos(\theta)\tanh(ax+\textit{i}k){\rm e}^{-\textit{i}a^2t},
    \end{align}
\end{subequations}
where $a,k,\theta\in \Re$, $k\neq\pi$ with,

The bright and dark soliton solutions given by
eqs.\eqref{brightone} and \eqref{darkone} are identical to the one
generated for PT symmetric scalar nonlocal NLS equation
\cite{aksarma} and the details of the derivation of the bright and
dark solitons are found in ref.\cite{liming}.

To understand the impact of nonlocal interaction, we need higher
order soliton solutions. For that, we employ Darboux
transformation approach to construct the two soliton solutions of
the following form:
\begin{equation}\label{soliton1}
    q_k^{(2)}=\alpha_k\left[\frac{\cosh(\zeta_1)\cosh(A-2 \textit{i}\eta_1)+\cos(\eta_1)\cosh(B+2\zeta_1)}{\cosh(\zeta_1)\cosh(A)+\cos(\eta_1)\cosh(B)}\right]{\rm e}^{\textit{i}\phi t}
\end{equation}
\begin{align}
       A=& X_{1} + X_{2} - \zeta_{1} + a_{1},\;\; B = - X_{1} + X_{2} + \textit{i} \eta_{1} - \textit{i} b_{1}, \notag\\
       X_{1}= & \textit{i} \sqrt{J_{1}} \sinh \left(\zeta_{1} + \textit{i} \eta_{1}\right)\left( x + \sqrt{J_{1}}\cosh\left(\zeta_{1}+\textit{i}\eta_{1}\right)t\right),\notag \\
       X_2=& \textit{-i} \sqrt{J_1} \sinh \left(\zeta_{1} - \textit{i} \eta_{1} \right)\left(-x + \sqrt{J_{1}}\cosh\left(\zeta_{1}-\textit{i} \eta_{1}\right)t\right), \notag\\
       J_1=& -\left(a\alpha_{1}^{2}+2 b\alpha_{1}\alpha_{2} + d\alpha_{2}^2\right),
\end{align}
where $\zeta_{1}$ and $\eta_{1}$ are spectral parameters and
$\alpha_1,\beta_1,a_1,b_1,a,b,d$ are real arbitrary parameters.

\begin{figure}[!ht]
\centering\includegraphics[width=0.5\linewidth]{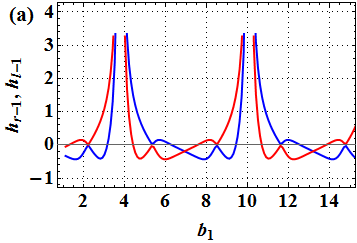}
\centering\includegraphics[width=0.5\linewidth]{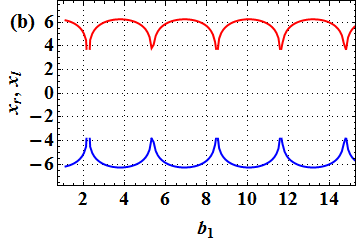}
\caption{(a) Blue and Red lines show the heights of the two
solitons $h_r$ and $h_l$, respectively,  with respect to the
uniform background of one. (b) The two solitons locations $x_r$
and $x_l$. }\label{2dim}
\end{figure}

\begin{figure}[!ht]
\centering\includegraphics[width=0.5\linewidth]{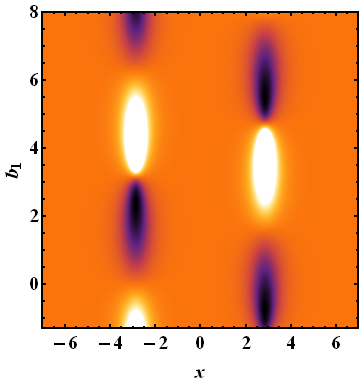}
\caption{Density plot of $|q_1|$ showing the conversion of
Dark-Dark soliton into Bright-Bright, Dark-Bright (or) Bright-Dark
by manipulating the parameter $b_1$. }\label{b1}
\end{figure}

\section{Family of exact solutions}

Equation~\eqref{soliton1} corresponds to a host of exact solitonic
solutions which can be obtained for specific values of the
parameters. By inspection, we found the group of solutions
containing: i) a pair of solitons, ii) a solitonic oscillatory
wave, iii) Akhmedeiv breather. All these solutions were on a
uniform background. Interestingly, we found that these solutions
exist also on a nonuniform background.

The pair of solitons solutions include Bright-Bright (BB),
Dark-Dark (DD), Bright-Dark (BD), and Dark-Bright (DB) solitons.
The main features of these solutions are obtained by noticing that
the denominator of Eq.~\eqref{soliton1} contains two peaks which
result in two soliton profiles. Therefore, the locations of the
two solitons are determined by the roots of the first derivative
of the denominator of Eq.~\eqref{soliton1}, namely
\begin{align}
x_{r}=&\Re\left[J_1^{-1/2}\cosh^{-1}\left(\frac{A_1}{A_2}\right)\csc(\eta_1)\right],\notag
\end{align}
where
\begin{align}
A_1=&\sqrt{\cos(\eta_1)\left(\cos(\eta_1)-\cos(b_1-\eta_1)\cosh(a_1)\right)}\notag\\
A_2=&\sqrt{1+\cos(2\eta_1)} \notag\\
x_l=&-x_{r}\notag
\end{align}
where $x_r$ and $x_l$ are the locations of the soliton on the
right and left of the origin. The corresponding amplitudes of the
two solitons, $h_r$ and $h_l$, are then calculated by substituting
these expressions back in $q_1$. The separation between the two
solitons and their peaks are thus determined by the parameters
$J_1$, $a_1$, $\eta_1$, and $b_1$. In Fig.\ref{grid}, we show the
whole family of independent solutions of Eq.~\eqref{soliton1}. In
addition to the pairs of soliton solutions, we show an oscillatory
solution, for $J_1<0$, identified as the Akhmedeiv breather
\cite{AB1,AB2} shown in Fig.\ref{grid}(e). Furthermore, we found
that the parameter $\zeta_1$ represents the separation between the
background levels between the two sides of the $x$-axis. This is
obtained by taking the limits $\lim_{x\rightarrow\pm\infty}|q_1|$
which will be $\exp{(2\zeta_1)}$ for the right side and
$\exp{(-2\zeta_1)}$ for the left side. Therefore the difference
between the two background levels equals
$\exp{(2\zeta_1)}-\exp{(-2\zeta_1)}$. This case is shown
Fig.\ref{grid}(f).

Remarkably, we found that all kinds of pair of soliton solutions
can be obtained by manipulating one parameter, namely $b_1$.
Fixing all other parameters, we plot in Fig.\ref{2dim} the
solitons locations and their peak values with respect to the
background. This figure gives the values of $b_1$ where we have
dark-bright ($b_1=2.8$), bright-dark ($b_1=4.3$), bright-bright
($b_1=3.9$) and dark-dark ($b_1=5.0$) solitons. The separation
between the two solitons is controlled most sensitively by $a_1$.

\begin{figure}[!ht]
\centering\includegraphics[width=0.8\linewidth]{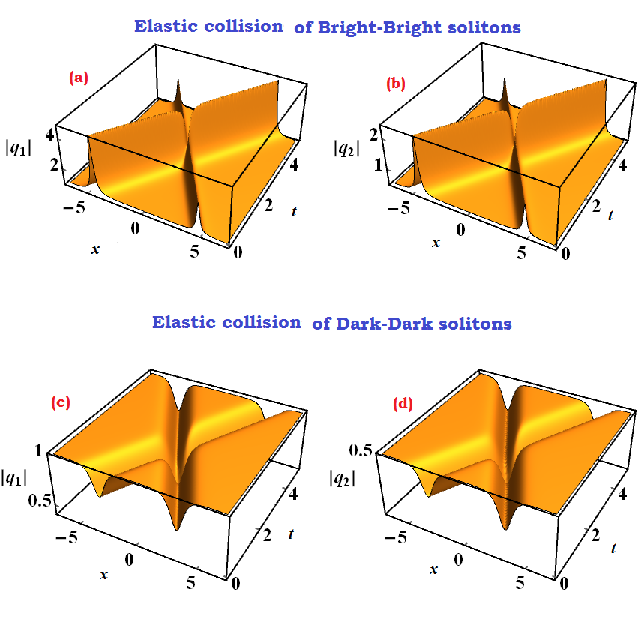}
\caption{Density plots of $\lvert q_1\rvert$ and $\lvert
q_2\rvert$ showing elastic collisions of (a,b) Bright and (c,d)
Dark solitons for the choice of parameters $\alpha_1=1,
\alpha_2=0.5, a=-4, b=-1, b=0, d=0, a_1=8, \zeta_1=0$ with
$\eta_1=0.14\pi, b_1= 3.8 $ and $\eta_1=0.25\pi, b_1=7$ for bright
and dark solitons respectively }\label{elas}
\end{figure}

\section{Collisional Dynamics of Solitons and their Conversion}
As we mentioned before, the classical (local) coupled nonlinear
Schr\"odinger equation admits only bright and dark soliton
solutions when their interatomic interaction were attractive or
repulsive, respectively. Unlike the usual behaviour, the study of
this parity-time symmetry in nonlocal integrable coupled NLS
equation surprisingly admits both bright and dark solitons for the
same nonlinearity. So, we step up our investigation and show that
one can convert the bright mode into a dark mode by selectively
finetuning the amplitude-dependent parameter in a two soliton
solution. In other words, one can convert the bright-bright
soliton into bright-dark or dark-dark by finetunig the
amplitude-dependent parameter without violating the integrability.
The conversion of soliton bound states with respect to parameter
$b_1$ at time $t=0$, is shown in Fig. \ref{b1}. The evolution of
the dark-dark soliton is as shown in Fig. \ref{grid}(a) for the
choice of parameter $b_1$=7. One can accomplish the conversion of
dark-dark soliton into bright-bright soliton as shown in
Fig.\ref{b1} and the corresponding evolution is as shown in
Fig.\ref{grid}(b) by tuning the amplitude-dependent parameter to
$b_1$=3.9. The choice of parameter $b_1$=2.8 and $b_1=5$ lead to
the conversion of bright-bright to dark-bright and bright-dark
solitons, respectively, as shown in Fig. \ref{b1} and the
respective evolution is shown in Fig.\ref{grid}(c) and (d)
respectively.

It is interesting to notice, as shown in Fig. \ref{grid}(f), that
in the case when the levels of the background are unequal, the
solitons reflect off the boundary between the two values of the
background. This is in contrast with the uniform background case
where the solitons transmit through each other and continue their
trajectory in a straight path.

Being an exact solution, the two soliton solution conserves the
energy. Therefore, solitons conversion between the different
members of the family of solutions (BB, DD, BD, and DB) is
guaranteed to conserve energy. We confirm this viewpoint by
showing the elastic collision of bright and dark solitons in
fig.\ref{elas} (a,b) and (c,d) respectively. It is only with
nonlocality that such combinations of solitons and their
conversion exists.

\section{Conclusion} In this paper, we investigated $\cal{PT}$ symmetric
CNLS equation employing Darboux transformation to generate a
family of exact solitonic solutions including a combination of
bright and dark solitons in addition to breathers both on a
uniform and nonuniform backgrounds. We have shown that one can
convert a bright bound state into a dark bound state in a mixed
soliton solution by manipulating the free (or arbitrary) parameter
associated with the system. We believe that the above phenomenon
occurs due to the nonlocal nature of the dynamical system. A
refinement on the present model would be incorporating $\cal{PT}$
symmetry through a potential while retaining the traditional cubic
nonlinear term. Such a model is more realistic and may be tested
experimentally, which we leave to future study.

\section{Acknowledgements:}
UAK and PSV acknowledge the support of UAE University through the
grant UAEU-UPAR(7) and UAEU-UPAR(4). RR wishes to acknowledge the
financial assistance received from Department of Atomic
Energy-National Board for Higher Mathematics (DAE-NBHM) (No.
NBHM/R.P.16/2014) and Council of Scientific and Industrial
Research (CSIR) (No. 03(1323)/14/EMR-II) for the financial support
in the form Major Research Projects. LL acknowledge the financial
support received from National Natural Science Foundation of China
(Contact No. 11401221). PSV and RR wish to acknowledge the
technical discussion with Prof. D-S. Wang, Beijing Information
Science and Technology University, Beijing, China.

\section{Appendix: Lax pair and Darboux transformation of the system given by eq.\eqref{eq:ptcnnls}}
For nonlinear partial differential equations, Darboux
transformation \cite{DT} is applied in an indirect manner. One
starts by finding a linear system of equations for an auxiliary
field ${\bf \Psi}$ in the form ${\bf \Psi}_x$ = ${\bf U. \Psi}$
and ${\bf \Psi}_t = {\bf V. \Psi}$, where ${\bf \Psi}$, ${\bf U}$
and ${\bf V}$ are $3 \times 3$ matrices. The pair of matrices
${\bf U}$ and ${\bf V}$, known as the Lax pair, are functionals of
the solution of the differential equation. The consistency
condition of the linear system, ${\bf \Psi}_{xt} = {\bf
\Psi}_{tx}$, should be equivalent to the differential equation.
The linear system and hence its consistency condition are
covariant under the Darboux transformation. Therefore, applying
the Darboux transformation on the linear system results in a new
consistency condition which is equivalent to a new differential
equation that is covariant with the old one. The new differential
equation is satisfied by a new solution. In the following two
subsections, we describe this procedure in a more detailed manner.

\subsection{Darboux transformation}

Consider the general form of nonlinear partial differential
equation
\begin{align}
F_{k}[Q(x,t),Q^{\ast}(x,t),Q_{t}(x,t),Q_{xx}(x,t)]=0,\;\;(
Q(x,t)=q_{k}(x,t)\; \& \; k=1,2 ).
\end{align}
The auxiliary field is represented by a $3 \times 3$  matrix:

\begin{align}\label{afield}
{\bf \Psi}=\begin{pmatrix}
\psi_1(x,t) & \psi_2(x,t) & \psi_3(x,t) \\
\phi_1(x,t) & \phi_2(x,t) & \phi_3(x,t) \\
\chi_1(x,t) & \chi_2(x,t) & \chi_3(x,t)
\end{pmatrix}.
\end{align}
The linear system of equations of the auxiliary field is formally
written as an expansion in powers of the constant eigenvalue
matrix
\begin{align}
{\bf \Lambda}& =
\begin{pmatrix}
\lambda_1 & 0 & 0 \\
0 & \lambda_2 & 0 \\
0 & 0 & \lambda_3
\end{pmatrix},
\end{align}
as follows
\begin{subequations}\label{laxcondition}
\begin{align}
{\bf \Psi}_{x}& = {\bf U_{0}}{\bf \Psi} + {\bf U_{1}}{\bf \Psi}
{\bf \Lambda},  \\
{\bf \Psi}_{t}& = {\bf V_{0}}{\bf \Psi} + {\bf V_{1}}{\bf \Psi}
{\bf \Lambda} + {\bf V_{2}}{\bf \Psi}{\bf \Lambda}^{2},
\end{align}
\end{subequations}
where,
\begin{align}
{\bf U_{0}}=
\begin{pmatrix}
0 & q_{1}(x,t) & q_{2}(x,t) \\
-r_{1}(x,t) & 0 & 0 \\
-r_{2}(x,t) & 0 & 0%
\end{pmatrix},
\;\;\; {\bf U_{1}}=
\begin{pmatrix}
1 & 0 & 0 \\
0 & -1 & 0 \\
0 & 0 & -1%
\end{pmatrix},\notag
\end{align}
\begin{align}
{\bf V_{0}}=\frac{\textit{i}}{2}
\begin{pmatrix}
q_{1}(x,t) r_{1}(x,t)+ q_{2}(x,t) r_{2}(x,t)  & q_{1x}(x,t) & q_{2x}(x,t) \\
r_{1x}(x,t) & -q_{1}(x,t) r_{1}(x,t) & -q_{2}(x,t) r_{1}(x,t) \\
r_{2x}(x,t) & -q_{1}(x,t) r_{2}(x,t) & -q_{2}(x,t) r_{2}(x,t)%
\end{pmatrix}, \notag
\end{align}
\begin{align}
{\bf V_{1}} &=
\begin{pmatrix}
0 & -q_{1}(x,t) & -q_{2}(x,t) \\
r_{1}(x,t) & 0 & 0 \\
r_{2}(x,t) & 0 & 0%
\end{pmatrix},
\;\;\; {\bf V_{2}}=\mathrm{i}
\begin{pmatrix}
1 & 0 & 0 \\
0 & -1 & 0 \\
0 & 0 & -1%
\end{pmatrix}, \notag
\end{align}
along with the transformation on the complex conjugates
\begin{align}
r_1(x,t) &= q_{1}^{\ast}(-x,t),
\notag\\
r_2(x,t) &= q_{2}^{\ast}(-x,t). \notag
\end{align}
The consistency condition ${\bf \Psi}_{xt}={\bf \Psi}_{tx}$ leads
to
\begin{subequations}\label{dtcont}
\begin{align}
&{\bf U_0}_t - {\bf V_0}_x + \Big[{\bf U_0},{\bf V_0}\Big]={\bf 0}, \\
&{\bf U_1}_t - {\bf V_1}_x + \Big[{\bf U_0},{\bf V_1}\Big] +
\Big[{\bf U_1},{\bf V_0}\Big] ={\bf 0},\\
&{\bf U_2}_t - {\bf V_2}_x + \Big[{\bf U_0},{\bf V_2}\Big] +
\Big[{\bf U_1},{\bf V_1}\Big] + \Big[{\bf U_2},{\bf V_0}\Big]={\bf 0},\\
&\Big[{\bf U_1},{\bf V_2}\Big] + \Big[{\bf U_2},{\bf V_1}\Big]={\bf 0},\\
&\Big[{\bf U_2},{\bf V_2}\Big]={\bf 0}.
\end{align}
\end{subequations}
These equations are obtained by equating the coefficients ${\bf
\Lambda^{0}}, {\bf \Lambda^{1}}, {\bf \Lambda^{2}}$ and ${\bf
\Lambda^{3}}$ in ${\bf \Psi}_{xt}$ to the corresponding ones in
${\bf \Psi}_{tx}$. The matrices ${\bf U_0}$, ${\bf V_0}$ are the
Lax pair of model equation \eqref{eq:ptcnnls} \& the consistency
condition, eq.\eqref{dtcont}(a) is equivalent to equations (6,7).

Considering the following version of Darboux transformation
\begin{align}\label{dteq}
{\bf \Psi}[1] = {\bf \Psi}  {\bf \Lambda} - {\bf \sigma \Psi},
\end{align}
where, ${\bf \Psi}[1]$ is the transformed field and ${\bf \sigma}
= {\bf\Psi_{0}} . {\bf \Lambda} . {\bf\Psi_0^{-1}} $. where  ${\bf
\Psi_0}$ is the known solution of the linear system given by
\eqref{laxcondition}. To determine the solution, the coefficients
of the linear system should be known explicitly. These
coefficients are functionals of the solution of the differential
equation $Q$. Thus, determining the coefficients of the linear
system requires knowing an exact solution of the differential
equation. This solution is known as the seed solution, which we
denote here by $Q_0$. It is in the very nature of the Darboux
transformation method that new exact solutions are only obtained
from other exact solutions. The transformed field ${\bf \Psi}[1]$
is required to be a solution of a linear system that is covariant
with the system \eqref{laxcondition}, namely
\begin{subequations}\label{laxconditionnew}
\begin{align}
{\bf \Psi}[1]_{x} &= {\bf U_{0}}[1]{\bf \Psi}[1] + {\bf
U_{1}}[1]{\bf \Psi}[1]{\bf \Lambda},  \\ \notag\\
{\bf \Psi[1]}_{t} &= {\bf V_{0}}[1]{\bf \Psi}[1]+{\bf
V_{1}}[1]{\bf \Psi}[1]{\bf \Lambda}+{\bf V_{2}}[1]{\bf \Psi}[1]
{\bf \Lambda}^{2}.
\end{align}
\end{subequations}
Requiring the system given by Eqs.\eqref{laxconditionnew} to be
covariant with the system of  Eqs. \eqref{laxcondition} leads to
the consistency condition
\begin{align}\label{neweqbydt}
{\bf U_0}[1]_t - {\bf V_0}[1]_x + \Big[{\bf U_0}[1],{\bf
V_0}[1]\Big]={\bf 0},
\end{align}
this is covariant with Eq. \eqref{laxcondition}(a). Similar to
model equation\eqref{eq:ptcnnls}, this new consistency condition
will be equivalent to a differential equation and it must be co
variant with equation\eqref{eq:ptcnnls}.
\begin{align}
{\bf U_0}[1]_t - {\bf V_0}[1]_x + \Big[{\bf U_0}[1],{\bf
V_0}[1]\Big]=\begin{pmatrix}
0 & F_{1} & F_{2} \\
F_{1}^{\ast} & 0 & 0 \\
F_{2}^{\ast} & 0 & 0%
\end{pmatrix}=0,
\end{align}
and the above new lax condition must be equivalent to the same
model equation under consideration.This means that $Q[1]$ is the
new solution of the same differential equation for which  $Q_0$ is
the seed solution.

To find ${\bf U_0}[1]$ or ${\bf V_0}[1]$ and hence $Q[1]$, we
substitute for ${\bf\Psi}[1]$ from eq.\eqref{dteq} in
eq.\eqref{laxconditionnew} using eqs.\eqref{laxcondition} and then
equate the $\Lambda$ powers to zero to get
\begin{align}
{\bf U_{0}}[1] &= {\bf \sigma} {\bf U_{0}} {\bf\sigma}^{-1} +
{\bf \sigma}_{x} {\bf \sigma}^{-1},\\
{\bf U_{0}}[1] &= {\bf U_{0}}+[{\bf U_1}, {\bf \sigma}].
\end{align}

The new solution $Q[1]$ can be calculated using either of these
two equations which can be shown to be equivalent. Notice that the
quantities on the right-hand side are calculated using the seed
solution $Q_0$.

To summarize, a nonlinear differential equation can be solved with
the Darboux transformation method by first finding an exact (seed)
solution, $Q_0$, to the differential equation and finding a linear
system for an auxiliary field ${\bf \Psi}$ that is associated with
the differential equation through its consistency condition. Using
the seed solution, a solution of the linear system, ${\bf
\Psi_{0}}$, is found. The linear system is then transformed into a
new one via the Darboux transformation. Thus, the coefficients of
the new linear system which are functionals of the new solution of
the differential equation, $Q[1]$ will be related to the
coefficients of the original linear system which are functionals
of $Q_0$. This relation gives the new solution $Q[1]$ in terms of
the seed solution $Q_0$. This procedure will be further
illustrated with our specific example in the following Section.

\subsection{Soliton solutions}
The values for the lax pair matrices ${\bf U_{0,1,2}}$ and ${\bf
V_{0,1,2}}$ are derived by trial and error method and their exact
forms are given above. In the following, we consider two cases,
namely zero and non-zero seed solutions. With zero seed, we obtain
the fundamental solitons while with the non-zero seed, we obtain
higher order solitons.

\subsubsection{Zero seed}

One can depart from two different paths to derive soliton
solutions in Darboux transformation. The first choice is to feed
the field variables $q_1(x,t)$ and $q_2(x,t)$ to be zero and use
lax pair matrices and adopt the procedure enumerated above to
obtain,

\begin{align}
\left(
\begin{array}{ccc}
 C_{11} e^{i \left(\lambda _1 x- \lambda _1^2 t\right)}  & C_{12} e^{i \left(\lambda _2 x-\lambda _2^2 t\right)}  & C_{13} e^{i \left(\lambda _3 x- \lambda
   _3^2 t \right)}  \\
 C_{14} e^{-i \left( \lambda _1 x- \lambda _1^2 t \right)}  & C_{15} e^{-i \left( \lambda _2 x- \lambda _2^2 t\right)}  & C_{16} e^{-i \left(\lambda _3 x-\lambda _3^2 t\right)}  \\
 C_{17} e^{-i \left( \lambda _1 x- \lambda _1^2 t \right)}  & C_{18} e^{-i \left( \lambda _2 x- \lambda _2^2 t\right)}  & C_{19} e^{-i \left(\lambda _3 x-
   \lambda _3^2 t \right)}  \\
\end{array}
\right),
\end{align}
which leads to the general one soliton solution of the following
form

\begin{equation}
q_{k}^{(1)}(x,t)=\frac{2 i
(C_{11}(A_{1k}(\lambda_1-\lambda_2)-A_{2k}(\lambda_1-\lambda_3))+A_{3k}(\lambda_2-\lambda_3))}{C_{13}
B_{1k}-C_{12}B_{2k}+C_{11}B_{3k}},
\end{equation}

where,
\begin{align}
A_{11} &=C_{12} C_{19} e^{2 i ((\lambda_1+\lambda_2) x +
\lambda_3^2 t)}, A_{21} =C_{13} C_{18} e^{2 i
((\lambda_1+\lambda_3) x + \lambda_2^2 t)}, A_{31} =C_{12}
C_{13}C_{17} e^{2 i
((\lambda_2+\lambda_3) x + \lambda_1^2 t)},  \notag\\\notag\\
B_{11} &=(C_{15}C_{17}-C_{14}C_{18}) e^{2 i (\lambda_3 x +
(\lambda_1^2+\lambda_2^2) t)} ,\; B_{21}
=(C_{16}C_{17}-C_{14}C_{19}) e^{2 i (\lambda_2 x +
(\lambda_1^2+\lambda_3^2) t)},\notag \\\notag\\
B_{31} &=(C_{16} C_{18}-C_{15} C_{19})e^{2 i (\lambda_1 x +
(\lambda_2^2+\lambda_3^2) t)},\; A_{12} =C_{12}C_{16} e^{2 i
((\lambda_1+\lambda_2) x + \lambda_3^2 t)},\notag\\\notag\\
A_{22} &=C_{13}C_{15} e^{2 i((\lambda_1+\lambda_3) x + \lambda_2^2
t)},\; A_{32} =C_{12}C_{13}C_{14} e^{2 i ((\lambda_2+\lambda_3) x
+ \lambda_1^2 t)},\notag\\\notag\\
B_{12} &=-B_{11},\; B_{22}=-B_{21},\;B_{32}=-B_{31}. \notag \\
\end{align}
Choosing the parameters $\lambda_2= \textit{i}\lambda_3$,
$\lambda_3=\textit{i} a$, $\lambda_1=-\textit{i} a$ with
$C_{11}=\textit{i}(-C_{15}C_{17}+C_{14}C_{18})/\sqrt{C_{15}^{2}+C_{18}^{2}}$,
$C_{13}=\textit{i}(-C_{16}C_{18}+C_{15}C_{19})/\sqrt{C_{15}^{2}+C_{18}^{2}}$,
$Re[\lambda_1]=Re[\lambda_2]=0$, $C_{15} \neq -C_{18}$,
$(-C_{15}C_{17}+C_{14}C_{18})\neq 0$,
$(-C_{16}C_{18}+C_{15}C_{19})\neq 0$ with
$C_{15}^{2}+C_{18}^{2}=1$, one only needs to choose the constants
suitably to derive the nonsingular bright solution given by
equation \eqref{brightone}, where $a, k$ are real arbitrary
parameters.

\subsubsection{Non-zero seed}

We depart from the zero seed and choose the seed  solution of the
following form
\begin{subequations}
\begin{align}
q_1(x,t) &= \alpha_1 e^{i \phi t} \\
q_2(x,t) &= \alpha_2 e^{i \phi t}
\end{align}
\end{subequations}
where,$\phi=a \alpha_1^2 + d \alpha_2^2 + 2 b \alpha_1 \alpha_2$
and $\alpha_1$, $\alpha_2$ $a,b,d$ are real free parameters and
follow the procedure mentioned above so that the fundamental
solution becomes
\begin{equation*}
    \Psi_i=JL_iE_i,\,\, J=\mathrm{diag}(1,\textit{e}^{\textit{i}\phi},\textit{e}^{\textit{i}\phi}),
\end{equation*}
where
\begin{equation*}
    L_i=\begin{bmatrix}
          1 & 1 &0\\[8pt]
          {\displaystyle \frac{\alpha_1}{\xi_{i,1}}} & {\displaystyle \frac{\alpha_1}{\xi_{i,2}}}& \beta_1 \\[10pt]
          {\displaystyle \frac{\alpha_2}{\xi_{i,1}}} & {\displaystyle \frac{\alpha_2}{\xi_{i,2}}}& \beta_2 \\
        \end{bmatrix},\,\, E_i=\mathrm{diag}\left(c_{i,1}\textit{e}^{\theta_{i,1}},c_{i,2}\textit{e}^{\theta_{i,2}},c_{i,3}\right)
\end{equation*}
and
\begin{equation*}
\begin{split}
\theta_{i,1}(x,t)=&\textit{i}\xi_{i,1}\left(x+\frac{1}{2}\xi_{i,1}t\right),\,\, \theta_{i,2}(x,t)=\textit{i}\xi_{i,2}\left(x+\frac{1}{2}\xi_{i,2}t\right),\\ \beta_1=&-(b\alpha_{{1}}+d\alpha_{{2}}-\textit{i}c\alpha_{{1}}),\,\, \beta_2=a\alpha_{{1}}+b\alpha_{{2}}+\textit{i}c\alpha_{{2}}, \\
\xi_{i,1}= &\lambda_i+\sqrt{\phi+\lambda_i^2},\,\,
\xi_{i,2}=\lambda_i-\sqrt{\phi+\lambda_i^2}.
\end{split}
\end{equation*}
Following the same procedure described above, one can obtain the
general soliton solution of the following form
\begin{equation}\label{soliton-formula}
    q_k[1]=\alpha_k\left[1+\frac{{\displaystyle \sum_{l=1}^{2}\sum_{m=1}^{2}\frac{c_{1,l}^*c_{1,m}{\rm e}^{\theta_{1,l}^*(-x,t)+\theta_{1,m}(x,t)}}{\xi_{1,m}}+c_{1,3}\beta_k\sum_{l=1}^{2}c_{1,l}^*{\rm e}^{\theta_{1,l}^*(-x,t)}}}{{\displaystyle -\sum_{l=1}^{2}\sum_{m=1}^{2}\frac{c_{1,l}^*c_{1,m}{\rm e}^{\theta_{1,l}^*(-x,t)+\theta_{1,m}(x,t)}}{\xi_{1,l}^*+\xi_{1,m}}+\frac{|c_{1,3}|^2B}{2(\lambda_1^*+\lambda_1)}}}\right]{\rm e}^{\textit{i}\phi t},
\end{equation}
where $B=a|\beta_1|^2+d|\beta_2|^2+(b+\textit{i}
c)\beta_1\beta_2^*+(b-\textit{i}c)\beta_1^*\beta_2.$ The above
generalized solution will reduce to several classes of solutions
depending upon the choice of parameters without affecting the
integrability and the validity of the solution to the
corresponding model equation under consideration. For the choice
of parameters $b=c=\alpha_1=a_1=0$ with limit $l=m=1$ gives us a
chance to derive several class of soliton solutions in which dark
solitons given by equation \eqref{darkone} is one among them. The
details of the derivation of dark solitons is given in detail in
Ref.\cite{liming}.

\subsubsection{Two-soliton solution}
Higher order solutions are then obtained by repeated actions of
the Darboux transformation in such a way that one obtains an
infinite chain of exact solutions. Repeating the same procedure
one more time by feeding the first order soliton as seed and
choosing the parameters $\phi<0$, $\lambda_1\equiv
\sqrt{-\phi}\cosh(\zeta_{1}+\textit{i}\eta_{1})$,
$\zeta_{1},\eta_{1} \in \mathrm{R}$, $\xi_{1,1}=\sqrt{-\phi}\;
\textit{e}^{\zeta_{1}+\textit{i}\eta_{1}}$,
$\xi_{1,2}=-\sqrt{-\phi}\;
\textit{e}^{-\zeta_{1}-\textit{i}\eta_{1}}$ and the arbitrary
constants $c_{1,3}=0$, $c_{1,1}c_{1,2}\neq 0$, then the solution
\eqref{soliton-formula} can be rewritten in a compact form as
given by equation \eqref{soliton1}.

\end{document}